# QUANTITY VERSUS QUALITY IN PUBLICATION ACTIVITY: KNOWLEDGE PRODUCTION AT THE REGIONAL LEVEL

Timur Gareev[1], Irina Peker[2*]

[1] Skolkovo Institute of Science and Technology, Bolshoy Boulevard 30, bld. 1, Moscow, 121205 Russia; https://orcid.org/0000-0002-6438-7772

[2] Immanuel Kant Baltic Federal University, A. Nevskogo Str. 14, Kaliningrad, 236016, Russia; https://orcid.org/0000-0002-5701-7538

*Corresponding author
Irina Peker
irinpeker@gmail.com

**Abstract**

This study contributes to the ongoing debate regarding the balance between quality and quantity in research productivity and publication activity. Using empirical regional knowledge production functions, we establish a significant correlation between R&D spending and research output, specifically publication productivity, while controlling for patenting activity and socioeconomic factors. Our focus is on the dilemma of research quantity versus quality, which is analysed in the context of regional thematic specialization using spatial lags.

When designing policies and making forecasts, it is important to consider the quality of research measured by established indicators. In this study, we examine the dual effect of research quality on publication activity. We identify two groups of quality factors: those related to the quality of journals and those related to the impact of publications. On average, these factors have different influences on quantitative measures. The quality of journals shows a negative relationship with quantity, indicating that as journal quality increases, the number of publications decreases. On the other hand, the impact of publications can be approximated by an inverse parabolic shape, with a positive decreasing slope within a common range of values. This duality in the relationship between quality factors and quantitative measures may explain some of the significant variations in conclusions found in the literature.

We compare several models that explore factors influencing publication activity using a balanced panel dataset of Russian regions from 2009 to 2021. Additionally, we propose a novel approach using thematic scientometric parameters as a special type of proximity measure between regions in thematic space. Incorporating spatial spillovers in thematic space allows us to account for potential cross-sectional dependence in regional data.

The article concludes with a discussion on the relevance and limitations of existing tools for modelling publication activity, along with some policy recommendations.

**Keywords** Publication activity, Research quality, Knowledge Production function, Panel data models, Scientometric tools, Russian regions

**Introduction**

The extensive body of literature on the structural and dynamic aspects of publication activity as a crucial component of knowledge production serves as a robust foundation for further research. In this paper, our main focus is the trade-off between the quality and quantity of publication production. This trade-off is exemplified by the debate presented in Butler (2003) and Van den Besselaar et al. (2017), which revolves around Australia's science policy implemented in the early 1990s, aiming to allocate funding based on publication output. This debate carries significant policy implications for countries striving to enhance their research and innovation efficiency. In this context, the case of Russia over the last decade serves as a valuable example of a country in a transitional phase, demonstrating the pursuit of convergence in research productivity. Recognizing the crucial role of knowledge productivity and its dependence on a research system's proximity to its production possibilities frontier, we have opted for the production function framework as a methodological approach for this study.

Butler (2003) focuses on relative indicators of scientific performance, such as the world shares of publications and citations in ISI. The author also used relative citation impact, which is the ratio of these shares (normally, this ratio should be around one). For most countries in Butler's comparison, increasing publication shares have been accompanied by an improved relative citation index (Butler, 2003: 147). Australia seemed to go against this trend, as its growth in research output appeared to come at the cost of impact. Butler states that "Australia's relative citation performance continued to slide because the journals which carried its articles were of lower impact" (Butler, 2003: 148). This effect is explained as a direct result of the performance appraisal scheme adopted by Australian universities during that period. Butler admits that this practice was new and affected the habits of researchers. To summarize, due to policy pressure, researchers on average tended to choose less impactful journals to produce more (and less good) papers.

This conclusion was sharply challenged by van den Besselaar et al. (2017). Following Martin (2017), we also hereafter refer to Butler's opponents as BHS. BHS note that in Australia, the average impact per publication has increased; the relative share of research output has improved without compromising quality, resulting in a significant overall increase in impact. Concisely, BHS claim that the new policy during the 1990s gave the Australian science system a new impulse, contributing to both higher productivity and higher quality in the long run. BHS posit that if output-based research funding affects research quality, "it is positive and not negative". Martin (2017) attempted to reconcile the two views, admitting that BHS somewhat overstated Butler's claims. In any case, we see a clash of two one-sided views on the problem, with somewhat opposite policy prescriptions.

We propose a model that contributes to the ongoing discussion on publication activity and research productivity, utilizing the regional knowledge production function (RKPF) and addressing many of the major claims made in the BNS - Butler debate. Our approach avoids country-level comparisons and instead employs sets of variables that are suitable for interpretation, including the field-weighted citation index (FWCI) and shares of publications in first or non-quartile journals. These indicators are relative, normalized, and widely recognized in modern bibliometric research. Our model integrates citation impact and incorporates journal impact through multiple variables, while also accounting for the interdisciplinary integration of data, which are challenging in original studies.

In the next section, we will examine Martin's clauses that address the nature of quality and impact. These clauses are points of agreement between both sides of the discussion. Ultimately, they reach the same conclusion that "… if one uses data on larger groups (teams, universities, or countries), citations are a fairly reliable and valid proxy for scholarly quality …, and this is the way we use citation impact in this paper − as did Butler ..." (van den Besselaar et al., 2017: 906).

Our findings reveal that aggregate measures of publication quality can be categorized into two groups: indicators of citation impact and journal prestige. These groups may have contrasting effects on quantitative publication counts, providing insight into the divergent views and contradictory conclusions in the current debate. Specifically, citation impact indicators, such as the field-weighted citation index (FWCI), show a positive relationship with publication counts, aligning with the argument put forth by BHS. However, we cannot disregard the stylized facts emphasized by Butler, as the share of Q1 publications exhibits a negative association with publication productivity. Additionally, the presence of 'non-classified' publications, associated with lower quality, contributes to formal productivity growth. We have verified these findings using publications data that includes and excludes conference proceedings, and our results remain consistent across both specifications.

Both countervailing effects have been identified as statistically significant, even when employing robust methodologies. Additionally, as an ancillary outcome of the study, we demonstrate that incorporating spatial lags into our version of the Regional Knowledge Production Function (RKPF) enhances model statistics and generally does not alter the main conclusions.

Our study goes beyond an acknowledgement of the valid points raised by both discussants. We propose a more coherent framework that captures countervailing forces and can disentangle the contributions of different factions within the research community.

Our work contributes to theory and practice, at least, in two ways: we estimate publication productivity on the regional level and focus on quality measures as explanatory variables, controlling for spatial spillovers in thematic space.

In the first part of the paper, we contextualize publication quality indexes within the framework of knowledge production function theory, explaining why the regional knowledge production function is suitable for examining potential trade-offs between quantity and quality, as well as its implications for the science system and possible policy interventions.

In the second part, we introduce the empirical model structure and explain how it is adapted to the given panel data to estimate two-way fixed effects models with spatial lags. We use a straightforward SLX type, following Halleck Vega & Elhorst (2015), and outline an original approach to estimating a *thematic proximity matrix* that captures spatial lags.

The third part of the paper presents a novel dataset on Russian regions from 2009-2021, which we have collected for this study, along with a discussion of some limitations of the data. We also use data on domestic patent applications as a control for alternative output competing for input resources (R&D expenditure per researcher).

In the fourth part, we compare different model specifications and report the main statistical results. Finally, we provide a conceptual discussion of the results, policy implications, and directions for further research to conclude the study.

# 1. Theoretical considerations: quality indexes and knowledge production paradigm

**Theoretical framework**

The 'production' of publications at the microlevel is characterized by a set of countervailing effects, including learning, experience, and network effects. We cannot directly match them with the effects of the learning curve, the experience curve, or network effects in the classical production theory because the latter require clearly defined notions of prices and costs (Levinsohn & Petrin, 2003). Nonetheless, we can understand the stylized mechanisms of those effects in the context of knowledge production.

Our position is that the process of publication constitutes a form of knowledge production and can, therefore, be characterized analytically through quality and quantity measures. Quantity measures have a lesser degree of ambiguity as an article in a peer-reviewed journal is typically recognized as a convenient unit of analysis. Although many adjustments and assumptions are required to render the number of articles a more homogeneous measure, the majority of these can be incorporated into the quality part of the model. The sole direct adjustment to the quantity measure that we make is utilizing the ratio of publications to the number of R&D employees.

The quality dimension is much more complex. The most challenging part is the multidisciplinary nature of the publications corpus. Even in separate fields of research (subject areas), the authors population is quite dispersed and unevenly distributed in terms of the value of their research contribution documented in publications. Conventionally, we measure this value indirectly through the impact of the research, calculated via citations in indexed journals.

Authors generally learn from their work through a process of knowledge accumulation. However, there are countervailing forces such as knowledge amortization and topic exhaustion (Candia et al., 2019). For instance, leading journals are often hesitant to publish similar results even from well-established authors. Nevertheless, more productive authors can switch between topics and build co-authorship networks to leverage their accumulated knowledge and experience. These networks may occupy limited space in top-rated journals, resulting in a trickle-down effect of knowledge dissemination in second-tier journals.

Information on journal reputations and academic conventions, such as the expected number and age of references per paper, is easily accessible to all authors at a minimal expense. Consequently, decision-making in the publication process follows a typical pattern. Authors representing their work aim to optimize their efforts and identify the most suitable journal for the article. In the normal course of action, authors evaluate their research piece and strive to submit it to a journal with a higher potential impact. If no suitable match is found, they pursue their second-best option. Substantial mismatches result in time delays, and responsible authors learn from these experiences.

Journal reputation serves as a crucial mechanism for signalling the quality of a publication at an international level. This, in turn, enables authors to establish their reputation, which is typically associated with the level of interest in their work. Moreover, citations serve as an indicator of the general interest in a topic, which helps authors adjust their publication strategy accordingly.

The above description represents a typical scenario of how knowledge production operates. However, policy interventions can distort the incentives for knowledge production.

If a policy designer has an oversimplified picture of the process, they may be satisfied with volumetric counts. This approach allows for simple country comparisons and claims that better publication dynamics reflect the growth of competitiveness or some other ultimate policy goal. However, this approach assumes a one-sided view of the effects and can lead to straightforward funding schemes that only encourage those who publish more in target sources.

This experience has been challenged by Butler. But, perhaps trying to sharpen her argument, she also eventually had drawn a one-sided picture. BHS in their critique have pointed out some other aspects like positive cumulative causation between production and impact but effectively also took a one-sided view of the problem.

**Quality indexes**

Scientometrics distinguishes between research productivity and research quality. The former describes the number of research publications, using different normalization variants: per researcher, per 1,000 population, etc. A traditional method for research quality assessment is peer review by colleagues working in the same field. There are also alternative research quality criteria built on bibliometric indicators, such as journal prestige (based on the impact factor) and citation accumulation (Seglen, 1997). Since impact factor calculation uses citation indicators, it is safe to assume that citation accumulation is a common measure of research contribution (Lindsey, 1989; Michalska-Smith & Allesina, 2017).

In other approaches, quality indicators are often based on journals' impact factors as a proxy for article quality, while citation is considered an indicator of impact (Haslam & Laham, 2010). Bornmann (2019) regards impact as an aspect of publication quality. In turn, impact can be assessed through both quality measures, such as peer review, and quantitative scientometric indicators. Such evaluation is necessary for measuring the impact of research on society (Sutherland et al., 2011).

In this article, we adopt the perspective of examining citation measures and the impact of research publications as qualitative indicators of knowledge production. Our focus is on understanding the relationship between these measures and the production of knowledge.

The discussion on the quality and number of publications has an immediate bearing on the search for an effective system of research funding. At the core of the debate is the question about the *connection between publication productivity and the quality of published findings.*

The introduction of a research funding system that uses direct quantitative measures of publication activity is linked by many researchers to the problem of the deteriorating quality of publications, which accompanies the increase in their number. This situation is in line with common sense since the focus on quantitative indicators encourages authors to publish in less influential periodicals to increase their publication count (Mathies et al., 2020; Butler, 2003).

Yet, some researchers strongly resent this approach, suggesting a positive correlation between the number of publications and citations (Lawani, 1986; Sandström & Van den Besselaar, 2016; Van den Besselaar et al., 2017). Therefore, a national policy to stimulate publication activity may urge researchers to publish their findings in top-quartile journals, which affects the number of publications (Bautista-Puig et al., 2021).

While admitting that the number of citations may increase as the number of publications grows, Binswanger (2015) emphasizes the growing number of substandard publications. And the increase in citations of such publications does not facilitate scientific progress. Kolesnikov et al. (2018), and Larivière & Costas (2016) also admit a positive relationship between the qualitative and quantitative indicators of publication activity. They also stress the need to take into account the effect on the indicators of the field of research, age, gender, work experience, the number of employees, culture, and institutional environment.

There is concern that citation-based funding may lead to a rise in self-citation per article (Abramo et al., 2021). This situation is a product of pressure from established research practices falling under the umbrella principle of "impact or perish" (Biagioli, 2016). This is, however, a difficult issue: there is evidence that the proportion of self-citations reduces with improving journal quality (Lawani, 1986). Moreover, contemporary analytical platforms, such as SciVal, offer the option of eliminating self-citations when calculating scientometric indicators.

The majority of researchers concur that R&D policy demands meticulous planning, as bibliometric indicators can have an instantaneous effect on the behaviour of researchers. Hasty introduction of clear-cut schemes entails systemic risks that manifest themselves differently in different countries. For instance, Baccini et al. (2019) describe the trend towards the increasing inwardness of Italy's national research system.

This debate suggests that the utilization of quantitative indicators to measure research productivity has a distinct impact on researchers' behaviour, as they respond promptly to changes in the incentive system. However, to the best of our knowledge, no comprehensive analysis has been conducted thus far to provide a systematic overview of the intricate relationship between qualitative and quantitative indicators of publication activity within a single model.

Within the knowledge production paradigm, it is essential to posit a causal relationship between quality and impact. However, a scientometric researcher cannot observe these variables directly (**Fig. 1**). The only observable variables are citations and journal rankings. Impact indicators, which should not be confused with impact as influence on knowledge production, typically rely on citations as their basis.

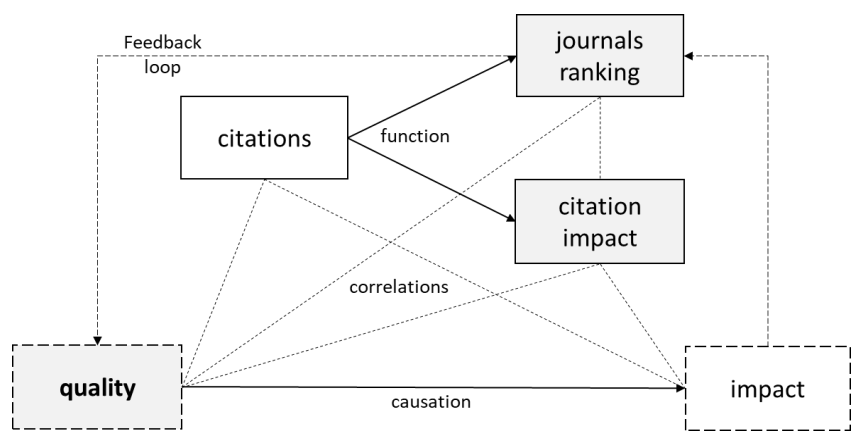

**Fig. 1** Conceptual framework for research quality at the aggregated level

Although journal rankings are inherently by-products of citation statistics, we prefer to see them as an independent set of variables because journals mediate the quality-impact relationship not only after, but most importantly, before a research piece becomes visible to the public. Journals play a crucial role in determining the visibility and impact of research, making them a crucial factor in assessing research quality[1].

---

[1] In our study, we collected and examined data on several potential quality indicators, such as citations per publication (CPP) and the share of publications with international collaboration (INTC), among others. We observed a high correlation among these indicators, particularly between CPP and INTC, as well as their correlation with the share of publications in Q1 journals. To address the issue of multicollinearity and ensure robust model estimation, we decided to reduce the number of explanatory variables in our models.

**Knowledge production functions**

A large corpus of the literature on publication activity[2] focuses on descriptive methods, usually analyzing trends in publications (or patents) or distributions of publications by topics or the social profile of authors, research collaboration and networking. Many sources attempt to model publication dynamics as a function of time, like Braun et al. (1989), Bettencourt et al. (2008), Zhou et. al. (2009) or Varun Shrivats & Bhattacharya (2014), to name just a few. Such models capture growth patterns and are probably suitable for forecasting. Yet, they often lack a specific theoretical background. In this regard, approaches based on the production function theory, or more specifically knowledge production, have gained wide recognition.

Griliches (1979) is a usual starting reference for knowledge production functions (KPF) as a theoretical concept. Also, Pakes & Griliches (1980) and Griliches (1990) discuss patents as an acceptable and accessible proxy for knowledge outputs and R&D expenses as reasonable inputs to a KPF applicable for empirical studies.

Later on, knowledge production functions have been extensively studied on different levels of aggregation – individuals, groups (collaborations and networks), organizations (firms and universities), regions, and countries. Unfortunately, the shift of KPF from an organizational to a regional level of *aggregation* is not theoretically straightforward due to a lack of understanding of the concept of knowledge production. Thus, the literature on RKPF often uses the term *empirical* models, emphasizing some deficiencies in their theoretical framework.

Among recent contributions to empirical RKPF, we draw on models and estimation approaches presented in **Annex I**. For instance, Kim & Lee (2015), and Vadia & Blankart (2021) propose regression models where the dependent variable is the number of publications (or its logarithm). Our modelling schemes are panel regression models of that kind. As **Annex I** shows, RKPF models often include spatial lags to model various interregional spillovers. In this regard, we highlight the contribution of Marrocu et al. (2013) emphasizing different spatial dimensions in knowledge production, including the institutional, technological, social and organizational proximities, along with geographical ones. Similar to our approach, they use R&D expenditure and human capital as the main internal inputs.

Other methodological approaches applied to Russian regional data can be found in Zemtsov & Kotsemir (2019) or Perret (2019), the former exemplifying the DEA methodology and the latter demonstrating quantile regression. Charlot et al. (2015) demonstrate a non-linear semiparametric approach to RKPF; Óhuallacháin & Leslie (2007) propose an extended set of explanatory variables.

Despite many relevant contributions, *quality measures* are not directly considered in the context of RKPF and innovation policy in general. This can be explained by most of the studies focusing on patenting activity, which is traditionally less equipped with quality measures than journal articles and conference proceedings.

**Theoretical implications for the empirical study**

Our literature review has revealed at least four issues in the research agenda deserving scholarly attention:
− ambiguity in the quality-quantity relations in publication production;

---

2 In what follows, we use the term *publication productivity* to refer to an indicator of publication *activity;* if no special comment is made, they are used interchangeably. In general, productivity is a relational indicator, most often, the ratio of output to input. Following Coelli et al. (2005), we distinguish between the two terms, with publication *efficiency* representing the closeness of publication productivity to its optimal levels.

– publication activity is a type of knowledge production, but previous studies have rarely looked at the interactions between patents and research articles as knowledge outputs;

– traditional spatial lags can hardly capture proximity in the thematic (disciplinary) space to estimate the level of research spillovers between different regions;

– expected temporal lags between resource deployment and knowledge output registration.

When trying to solve all the above issues at once, seemingly straightforward RKPF models quickly become demanding in terms of modelling tools. In the next section, we will try to use a balanced set-up to keep the models tractable and capable to address the key research question of our study, i.e., how the chase for publication scores influences the quality (research value) of the results.

**We are going to test the following hypothesis:**

– an increase in R&D expenses (in bulk and per researcher measures) is positively related to publication activity;

– in general, the influence of quality on volumetric measures is controversial since countervailing forces are in action;

– there should exist a spatial correlation in publication activities on the aggregate level, which can be captured by proximity in the thematic space.

The first hypothesis is technical in the sense that positive relations between productivity and resources are in agreement with both theory and common sense. This positive relationship has been demonstrated in many studies and serves as a litmus test of data and model validity.

The next crucial point in our study revolves around the classification of quality indicators into two main groups: those associated with journal quality and those linked to publication impact. For each group, we will select relevant indicators. Our aim is to examine the relationship between these quality indicators and productivity. If all significant quality indicators exhibit positive relationships with productivity, it would provide support for the critique put forth by BHS. On the other hand, if we observe negative relationships, it would generally support Butler's original position. However, based on our comprehensive theoretical review, we anticipate a more nuanced and intricate pattern in the data, suggesting that the relationship between quality indicators and productivity is likely to be multifaceted.

Finally, we are going to test our model in the heterogeneous environment of regions and aggregate publication activities across all thematic fields. It is reasonable to think that there should be some form of spatial correlation in the thematic space between regions, but we have not identified established approaches to this question in the existing literature. Thus, we are going to develop and test a relevant thematic proximity methodology.

**2. Empirical models**

In general form, KPF can be expressed as follows:

$$Knowledge\ output = f(R\&D\ expenses; R\&D\ employes, Knowlegde\ endowment)$$

On the regional level, knowledge output is represented by a variety of codified forms. Patents (or patent applications) are the most widely used dependent variable, although some authors experiment with the number of publications as well.[3] Principally, it is possible to model multiproduct output. That can be done using data envelopment analysis (see, for example, Chen et al., 2018 for discussion on DEA applications), but this family of models still lacks statistical properties and is sensitive to data quality.

To properly account for the influence of multiproduct cases in the framework of regression analysis, it is important to control for other knowledge outputs in addition to citation counts. These may include the number of patents or publications in national-language journals, among other factors.[4]

Publication quality is a less clear concept. However, if we assume the production function of the form: $e^q y = f(R)$, where $q$ stands for some quality index of the output and $R$ for a bundle of resources, after log transformation, it will result in: $lny = lnf(R) - q$. In this case, the minus sign before $q$ indicates that the quality index acts as a substitute for the quantitative measure.

As has been shown (**Annex I**), the literature provides a wide range of different regressors for RKPF (see Fritsch, 2002; Ramani et al., 2008; Fritsch & Slavtchev, 2010; Kim & Lee, 2015; Vadia & Blanckart, 2021).

The empirical model we use is based on the Cobb-Douglas form and includes a traditional set of control variables. This approach was chosen to facilitate replication and testing on alternative datasets.

$$Y = (Z, X)\beta + X_W \theta + \mu_r + \tau_t + \epsilon,$$

where $X$ are main explanatory variables; $X_W$ is a set of spatially lagged variables $X$ weighted with matrix $W$; $\mu_r$ and $\tau_t$ are region-specific and time-specific dummies, respectively.

As can be seen, the model is a spatial panel regression similar to the SLX type (for extensive discussion, see Baltagi, 2015; Halleck Vega & Elhorst, 2015).

One of the novel parts in this specification is our treatment of W. We propose to control for spillover effects using the concept of *thematic proximity of regions*. The term *thematic spatial lag* is used hereafter as well. Hence, each non-negative element of the matrix W indicates the intensity of the relationship between regions in academic fields.

To construct the thematic spatial lag matrix, we utilize a matrix that contains the distribution of publications across all subject areas (fields) in the Scopus database for each region. The shares of subject areas provide a thematic profile for each region. Thus, we have matrix $M$ of the size $n \times s$, where $n$ is the number of regions and $s$ is the number of subject fields in the Scopus database. Using matrix $M$ we get the correlation matrix of the size $n \times n$. By construction, a correlation matrix is symmetric and reflects similarities between regions in the thematic space.

We employ conventional methods to derive the spatial weights matrix $W$ from the correlation matrix to model proximity relations in the thematic space. As convention dictates, all diagonal elements of the spatial weights matrix are assigned a value of zero, and the rows are standardized to ensure they sum up to one. So, after standardization, each row still corresponds to a region and the elements of each row show positive spillovers from other regions to the given one[5].

---

[3] A stylized picture in economic theory postulates the difference between knowledge and technology as types of goods. While knowledge — at least, scientific — is regarded as a public good (with both non-rivalry and non-excludability properties), technologies are characterized as non-rival, but partially excludable. In literature, this difference often gets fuzzy (take RKPF as a notable example, where knowledge production is described with patents traditionally considered as a proxy for technologies).

[4] We acknowledge that the relationship between patents and publications is still subject to debate and remains unclear, even at the theoretical level. The question of whether they act as substitutes or complements is yet to be fully resolved. In our analysis, we have chosen to treat patents as control variables to account for their potential influence on research productivity. For further insights on this topic, we refer readers to the work of Kim & Lee (2015), who have provided recent discussions on this issue.

Eventually, one of our main empirical RKPF specifications gets the following form:

$$\begin{aligned}log(PUB21EMP_{r,t+1}) &= \beta_1 log(EXPEMP10_{rt}) + \beta_2 log(GDPCAP10_{rt}) + \beta_3 log(PAPEMP_{rt}) + \beta_4 FWCI_{rt} \\ &+ \beta_5 FWCI_{rt}^2 + \beta_6 Q1SH_{rt} + \beta_7 NQSH_{rt} + \theta_1 slFWCI_{rt} + \theta_2 slQ1SH_{rt} + \theta_3 slNQSH_{rt} \\ &+ \gamma_1 D_r + \gamma_2 D_t + \epsilon_{rt}\end{aligned}$$

where $D_r$ and $D_t$ stand for regional and time dummies, and the "sl" prefix reflects spatial lags of the respective variables. **Table 1** shows variables from the main empirical model.

**Table 1** Description of variables from the main empirical model

| Variable | Meaning | Status |
|---|---|---|
| log(PUB21EMP) | Log of the number of publications in Scopus journals per R&D employee, 2010-2021; hence, dependent variable values get shifted one period ahead of explanatory variables to capture a one-year average lag between efforts and article publication | dependent variable |
| log(EXPEMP10) | Log of real current R&D expenses per R&D employee, in 2010 prices | Control |
| log(GRPCAP10) | Log of real GRP (Gross Regional Product) per capita, in 2010 prices | Control |
| log(PAPEMP) | Log of a 3-year average of patent application per R&D employee 2009-2020; i.e., value for 2009 is calculated as the average of values in 2007, 2008 and 2009 weighted by 1, 2, 3 accordingly | Control |
| FWCI | Field-weighted citation impact | variable of interest |
| Q1SH | Share of publications in Q1 (First quartile) of Scopus journals | variable of interest |
| NQSH | Share of publications with no quartile in Scopus journals | variable of interest |

Note: more details about the variables and their data sources are discussed in the data section.

We collected mostly those explanatory variables that can be obtained for most modern states to make the empirical model less case-specific. Notice that the quality measures, $FWCI$, $Q1SH$ and $NQSH$, are not log-transformed.

The traditional dynamic panel approach suggests including temporal lags of dependent variables. However, we avoid dynamic modelling due to the small number of periods and, most importantly, due to the expected upward trend in the data due to the convergence phase in publication activity. With a longer dataset, the dynamic specification could be tested, as has been done in Kim & Lee (2015). Nevertheless, we introduce some quasi-dynamics by shifting the dependent variable one period ahead. This approach also helps to reduce endogeneity bias, which can be caused by temporal effects in the data.

**3. Data sources**

This study used research publications indexed in Russian regions from 2010 to 2021 as a proxy for knowledge outputs. Research publications, as understood in this article, include journal articles, reviews, preprints, conference proceedings, books, chapters, and other types of scholarly outputs. The Scopus database by Elsevier was chosen as the source of bibliometric data. We chose Scopus because it better represents developing countries and emerging economies, and in general, has a wider distribution of quality among its corpus compared with Web of Science. The data were collected in February-March 2020 (data on publications from 2020-2021 were augmented and corrected in 2022). Our data cover 78 Russian regions and federal cities.[6]

---

[5] The reasons why we may set all negative elements in our matrix to zero are as follows. Indeed, the thematic similarity of regions means that researchers within a region collaborate through many channels. These channels may include joint projects, monitoring profile publications, competing for research grants and industrial contracts, attending common conferences, academic mobility, and more. Research centres working in the same field influence each other even if they are not real collaborators; they do this via cooperation-competition mechanisms typical of the research community in general. At the same time, research centres that are negatively correlated in the thematic space are, in fact, neutral to each other in real life (which is best reflected by 0 spatial correlation). In other words, a representative researcher from natural sciences most likely ignores activities in humanities, and vice versa.

Since the metadata of publications contain information on the affiliation of each author, it was possible to distribute the publication dataset by region. Scopus does not offer the option of filtering articles at the regional level. Therefore, we had to do the following to attribute an article to a certain region:
− combine publications by organisations with a validated organisation profile in Scopus into publication sets for each Russian region according to the official administrative division.
− search for organisations without a validated profile in Scopus and add them to regional publication sets.

Along with metadata, Scopus contains information on different citation indicators and journal quality. The quantitative information on research publications was obtained from the SciVal analytical system. SciVal uses the standardized Snowball Metrics to conduct a comparative analysis of units at different administrative levels. We selected the following indicators of regional publication activity.

- **The number of publications in 2010-2021:** an indicator of the number of research contributions of all types over the study period and in each year. Since we compare the productivity of differently sized units, these indicators will be normalized by the number of R&D employees;

- **The field-weighted citation impact (FWCI) of publications from 2009-2021:** an indicator of citation, weighted by research field. It is calculated as the ratio of citations to the expected global average for the research area, type and publication year (Elsevier, 2020). If the FWCI is 1, the indicator of citation for contributions in questions corresponds exactly to the global average. For example, an FWCI of 1.30 means the level of citation of 30 percent is above the expected value. To account for the possible nonlinear behaviour of the FWCI variable in the model, we include a quadratic term for FWCI in the analysis.

- **Share of publications in Q1 (First quartile) of Scopus sources by the highest CiteScore rank in 2009-2021:** a quantitative assessment of the citation level of reviewed publications in periodicals. A Scopus-indexed journal is assigned to a quartile based on three bibliometric indicators: CiteScore, SNIP, and SJR. This is a qualitative measure of references to articles published in study journals in previous years. Within each thematic category, journals are ranked according to the CiteScore, SNIP, or SJR values and assigned to a certain quartile, from Q1 (the highest) to Q4 (the lowest). CiteScore is a Scopus metric showing the relative ranking of a journal in its field of knowledge. Q1 periodicals are among the 25 percent of the most cited in their research area.

- **Share of publications in non-quartile Scopus sources in 2009-2021** is a quantitative measure of the citations received by reviewed publications in periodicals. Non-Quartile sources may include recently indexed sources or Book Series and Conference Proceedings, for which a CiteScore rank has not been assigned. The most common methods for calculating publications number are full counting and fractionalized counting, as described by Sivertsen et al. (2019). In this study, a publication co-authored by researchers from different regions was attributed to each region.

---

[6] The Khanty-Mansi and Yamal-Nenets autonomous districts were amalgamated with the Tyumen region; the Nenets autonomous district with the Arkhangelsk region; the Jewish and Chukotka autonomous districts as well as Crimea and Sevastopol were excluded from the analysis due to data scarcity.

As control variables, we employed the following measures, which can be found in the annual statistical digest *Russian Regions. Socioeconomic Indicators* (Rosstat, 2021):

• **Regional R&D expenses per R&D employee in 2009-2020:** expenditure on R&D carried out by organizations using their own resources, without capital investment per R&R employee;

• **Gross Regional Product (GRP) per capita in 2009-2020**: the indicator was used as a control variable summarizing regional economic activity;

• **Number of patent applications per R&D employee in 2009-2020:** an indicator of patent activity in Russian regions, according to Rospatent (2021).

We used consumer price indexes for 2010-2020 from the same source to adjust monetary variables. Some of the indicators are normalized by the number of R&D employees, including researchers, technicians, and supporting staff. Overall, the direction of dependencies in models does not change when technicians and supporting staff are excluded from the calculation.

**Data limitations**

Measuring knowledge endowment and the spatial distribution of research facilities is a challenging task (due to limitations of available statistics and, more importantly, the fact that we study only the civil part of research activities). Thus, we assume that such capital endowment differences are captured by regional-specific fixed effects. In addition, we control for general economic indicators, i.e., gross regional product (GRP) per capita. This control also provides information about the relative economic size of regions.

At the same time, academic publications as a basis for measuring *research productivity* have advantages compared to traditional economic outputs due to their better incorporation of quality indicators. Besides, one may admit the definitive role of human capital (represented by R&D employees) in knowledge production. In practice, this means that the number of active researchers and R&D funding should be good proxies for knowledge production inputs.

In the literature, patents have long been acknowledged as a proxy for knowledge production, and, to the best of our knowledge, this topic has been elaborated in more detail in the economic literature due to its direct connection to the problem of technological development. Hausman et al. (1984) and Moed et al. (2005) have made notable contributions to the discussion on the relationship between patents and R&D. Additionally, Fritsch (2002) and Fritsch & Slavtchev (2010) have provided fruitful insights in this area of research. Publications share some similarities with patents, which allows us to adapt modelling approaches used for patents to publications. For example, both publications and patents have a temporal lag between the period of resource usage and the period of actual registration (as explained in **Annex I**). As a simplifying assumption, this lag is often set to one year, which is a reasonable and convenient approximation for the average publication. Fortunately, publication statistics are usually available one year ahead of regional economic indicators, allowing us to use the latest data to test the model. Additionally, the impact of patent activity on publication productivity can be complex and controversial, so we have collected data on patent applications to control for a possible crowding out effect. Although this relationship is not at the centre of our study, the modelling results may reveal interesting facts for further discussion.

One of the main limitations of regional studies is spatial correlation. This consideration calls for correcting the models by some proximity measures (e.g., in the form of additional controls). We collected data on the *thematic distribution* of publications in Russian regions in 2019 and modelled the proximity relations in thematic space as described in Section 2. We assume that the thematic profile is relatively stable during the period of study (about a decade).

### 4. Results

**Descriptive statistics**

After dropping and combining some regions (see Annex II), we have a *balanced* "short" panel with $N > T$, where $N = 78$ and $T = 12$ (the excluded regions contribute small volumes to the total publication activity, but the main reason is the missing data; we have kept as much observation as possible even though some regions have very low scores).

**Table 2** shows the descriptive statistics of the resulting dataset. For convenience, the table provides information for initial values along with log-transformed variables used in model estimations. The descriptive statistics suggest that publication productivity and resources are unevenly distributed across regions.

**Table 2** Descriptive statistics of variables in the main model (ratios and their logarithms)

|  | Units | Min | 1st Qu | Median | Mean | 3rd Qu | Max |
|---|---|---|---|---|---|---|---|
| PUB21EMP | Units per employee | 0.01 | 0.07 | 0.13 | 0.2 | 0.26 | 10.30 |
| EXPEMP10 | Th. Rub. per employee | 0.12 | 0.48 | 0.64 | 0.66 | 0.81 | 2.16 |
| GRPCAP10 | Rub. per capita | 48239 | 155447 | 213029 | 259290 | 285386 | 1584591 |
| PAPEMP | Units per employee | 0 | 0.03 | 0.05 | 0.08 | 0.08 | 1.06 |
| log(EXPEMP10) | - | -2.09 | -0.74 | -0.45 | -0.49 | -0.21 | 0.77 |
| log(GRPCAP10) | - | 10.78 | 11.95 | 12.27 | 12.30 | 12.56 | 14.28 |
| log(PAPEMP) | - | -7.56 | -3.57 | -3.05 | -3.01 | -2.49 | 0.06 |
| FWCI | - | 0.0 | 0.36 | 0.53 | 0.58 | 0.71 | 8.04 |
| Q1SH | % | 0.0 | 6.25 | 9.84 | 11.80 | 15.06 | 75.27 |
| NQSH | % | 0.0 | 8.02 | 12.95 | 15.21 | 19.06 | 100.0 |

Source: authors' estimations

A visual examination of the publication corpus indicates a rapid rise in publication activity between 2009 and 2020 (**Fig. 2**). Subjectively, we may observe *a sigmoid pattern* in the overall dynamics, although the duration of the time period is still relatively short to draw definitive conclusions based solely on the time series data. However, it is evident that the research system has been functioning below its potential productivity level. Notably, there has been a decrease in the recorded number of R&D employees during the study period, accompanied by a slight increase in real R&D spending per employee as a consequence of this reduction.

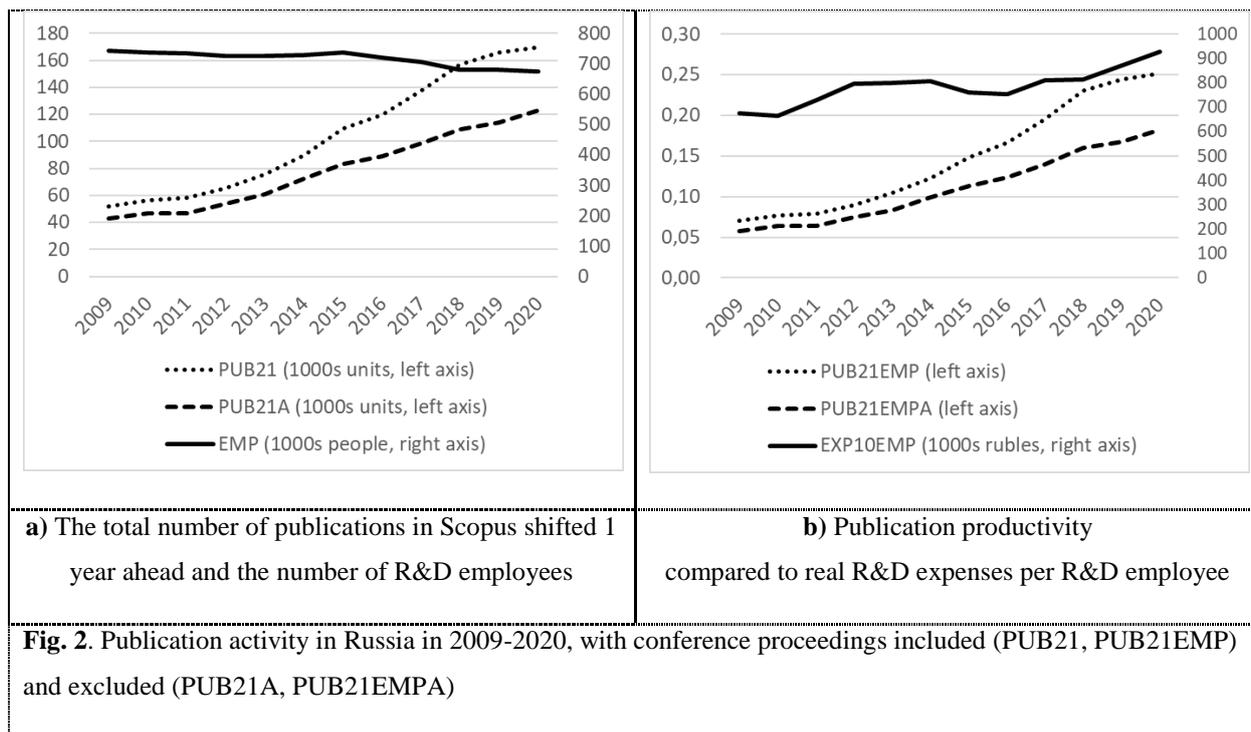

| a) The total number of publications in Scopus shifted 1 year ahead and the number of R&D employees | b) Publication productivity compared to real R&D expenses per R&D employee |

**Fig. 2**. Publication activity in Russia in 2009-2020, with conference proceedings included (PUB21, PUB21EMP) and excluded (PUB21A, PUB21EMPA)

Source: calculated by the authors

Dynamics in the full publication corpus and publications including Articles and Reviews differ only after 2018. We see the gap between the two variables is narrowing, which might be an indicator of stabilization and the convergence phase in research production. We emphasize that research productivity in Russia is still far from its steady state. In the example, we observe a simultaneous decline in the officially reported number of R&D employees and a rapid increase in volumetric terms (though the quality of research indicators is still comparatively low).

**Model estimations**

Our estimation strategy is as follows. We analyze two datasets: the main one encompassing all publications and the auxiliary one including only Articles and Reviews (approximately 70 percent of all publications). As mentioned, the main corpus of publications is more applicable to the study as it corresponds to the policy targets and performance indicators of the majority of research institutions.

The auxiliary dataset helps to verify general conclusions and identify possible flaws in the chosen approach to gain new insights.

For each regression set, we compare seven model versions with the following notations (**Table 3**).

**Table 3** Notations used for reporting model results

| Notation | Meaning | Explanation |
|---|---|---|
| ols / fe | Pooling / Fixed Effects | The pooling model usually shows a naïve approach to be contrasted with |
| ow / tw | One-way / Two-way | Using a dummy variable to control for one or two dimensions of the panel (one-way accounts only for region-specific effects) |
| Q | Quality variables included | Some or all quality variables are included in the model |
| Sl | Spatial lag | Inclusion of spatial lags for the quality variables presented in the model |
| A | Article and Reviews data | Using the auxiliary dataset with Articles and Reviews only (in general, all types of indexed publications are included) |
| non / noq | No quartile share / First | For quality variables, either NQSH or Q1SH variable is |

| | quartile share excluded | omitted to see the effect of the other one |

Most calculations are done with open-source R packages to ease potential replication and file sharing (mostly with "plm" package of Croissant & Millo, 2008; Millo, 2017).

We have consequently rejected pooled and random effect (RE) models for our data, but the result for pooled OLS as a reference point are still reported for chosen specifications.

For the main model, we present the results with robust standard errors (**Table 4**) as they provide a more accurate estimation and, in certain cases, offer clearer insights. We provide estimates and information criteria for the time-dummy factors in the model. This allows us to examine the effects and significance of these factors on the model's results. Auxiliary modelling results are presented in **Annex II** in a condensed format. Given the above notation, our main model is "fe.tw.q.sl", which is shown in the rightmost column of the regression tables.

First, our controls characterizing the research environment are significant in all settings and have positive signs, as expected. Interestingly, GRP per capita is not significant with robust errors. Therefore, its inclusion in models with fixed effects is merely a matter of tradition.

Our main attention is on the quality variables. FWCI and its squared terms are significant and consistent in all settings, strongly supporting the postulated positive relationship between quality and production. However, the way FWCI is constructed (centred around 1) suggests that the measure resists growth in the range of values higher than one. Therefore, a squared term is required in all models that include FWCI. Overall, an increase in productivity is associated with higher citation indexes and diminishing marginal growth.

Both Q1SH and NQSH have significant and expected signs. Publication productivity decreases with the growing share of high-quality publications, and, on the contrary, it increases with the growing share of publications of dubious quality. Other things being equal, our main model suggests that a 1 percent increase in the share of Q1 publications reduces, on average, the volumetric productivity by 1 percent, and by an additional 2 percent due to the spatial effect of quality growth. NQSH has the same magnitude of effects but in the opposite direction.

Although the correlation between Q1SH and NQSH is relatively low (-0.3), both variables are vying for inclusion in the model as their spatial lags show significance in all robust estimations.

While the spatial lag coefficients may not be significant in the main models with normal errors, the AIC tests incorporating spatial lags have shown to be the most effective among the selected specifications. Interestingly, spatial coefficients become significant (at different p-values) with robust standard errors. Moreover, including the spatial lags in the model makes time dummies less important, which allows admitting the spatial lag explanatory power. In any case, we recommend including them in the models of a regional level of aggregation.

An intriguing fact is that only Q1SHA loses its robustness in the full auxiliary dataset (**Annex II**).

If we exclude the year 2021 from the analysis, the main regularities for Q1SHA revealed in the main dataset remain in place. At the same time, NQSHA is not robust in the auxiliary dataset excluding the year 2021, although the coefficient has the expected sign.

There are several possible explanations for this fact (for example, the 2021 Scopus rules on excluding periodicals known for their controversial editorial policy or the short-term effects of the pandemic), but this may as well be an early sign of publication quality stabilization. It seems logical that this process is more visible in the Articles and Reviews dataset. We may evidence some stabilization of the quality of publication activity starting from 2020.

As mentioned earlier, quality variables in the models are lagging compared to the publication time as regards reducing endogeneity risks (due to a possible simultaneity bias). We have also tested models with quality and publication variables calculated without lag or as weighted averages of lagged and non-lagged panels. In both cases, the found regularities stay the same. We do not report the results here, but an interested reader can easily repeat the exercise with the data provided in the supplemental materials.

As a general conclusion for this part, we have found support for our hypotheses.

In general, models with two-way fixed effects have satisfactory explanatory power of around 0.8. Similar results are typical of many studies that we have mentioned in the article. The model demonstrates a strong increase in explanatory power when both fixed effects are added.

Regarding our main hypotheses, the following observations are relevant:

1) As expected, there is a significant positive relationship between publication productivity and resource inputs in the form of R&D spending per R&D employee. The elasticity is around 0.5, comparable with estimates found in the literature (e.g., Fritsch & Slavtchev, 2007: 243).

2) The model demonstrates that the coefficients of chosen quality indicators are statistically significant at least at the level of $p < 0.1$, even with robust statistics. (**Table 4** and **Annex II**).

The most striking result is their opposite relationship to publication productivity. Notice that the signs of quality variables are persistent among different specifications, even when not significant; that makes us more confident in the observed behaviour of the coefficients.

Including the square of FWCI has shown a strong and persistent negative sign of the quadratic term, which means that FWCI and publication productivity relation is inverse parabolic. At the same time, the sign of Q1SH is also always negative, while NQSH is positive.

3) Spatial thematic lags act in the same direction as the main variables, and their inclusion in the model improves their explanatory power.

**5. Discussion and conclusions**

Our literature review has revealed some ambiguity about the quality versus quantity dilemma. However, the model results go beyond our initial expectations. We expected that results would be unstable between specifications, but they demonstrated persistent but opposite behaviour.

Our initial guesses supported the stylized picture popularized by Butler (2003). Perhaps, we have underestimated some other effects documented in BHS.

We have found that quality indicators can be divided into two main groups, namely those related to publications impact and journals quality. Including all similar quality indicators is unnecessary: due to their collinearity, the coefficients of all but two quality indices were not significant. So, finally, only FWCI (publication impact), Q1SH and NQSH (journal quality) stayed in the model. Notice that we have decided to reflect journal quality from the top and bottom lines to grasp opposite behavioural patterns in the research community.

The negative quadratic term of FWCI deserves special attention. In practice, it looks very reasonable that in the range of extremely high values of FWCI (e.g., larger than 1.5), further increase in quality becomes more costly and cannot be supported without a decrease in volumetric terms. But this inverse parabolic relationship also means that we have a concavity in productivity under the most reasonable range of FWCI values. So, despite the overall positive relationship between FWCI and publication productivity, it demonstrates a kind of diminishing returns to quality increase.

This supports the general conclusion that the higher the quality of a research system, the more resistance it faces in terms of volumetric increase. This conclusion is completely in line with the Q1/NQ quality indicators.

Another important consideration about publications is their interdisciplinary mix. We have found some thematic spatial lags in explanatory variables in our data. Since there are many possible ways how these considerations can be integrated into a model, we have chosen in this study the simplest versions with spatial lags only.

We have concluded that neither BHS nor Butler's findings can be rejected because there is a negative relationship between the share of Q1 publications and publication productivity, and a positive relationship between the share of non-quartile publications, which suggests a degradation in quality.

Our results do not fully support Butler's view because she has focused on citation impact indicators, not the quality of journals. Nonetheless, we are fully aware that our dataset and approach are quite different, and we can only draw analogies between quality variables used in the two studies.

Our interpretation of the results from our model is as follows. There are countervailing forces in play, so it is highly unlikely that the overall effect on indicators will be one-sided, as both parties revealed. The researchers are not homogeneous in their capabilities, so for most productive authors performance-based funding is beneficial. They enjoy cumulative causation of quantity and impact, which is in line with the BHS position. Newcomers (especially those who start from scratch) enter the game under performance appraisal pressure and usually submit to less ambitious journals. The broadest representative group will in general follow the pattern suggested by Butler (2003). Performance appraisal usually requires results in each reporting period (usually year). For those who have a small number of simultaneously elaborated research pieces, it might be a wiser strategy to submit to a less impactful journal for a faster publication track. The same should be true for authors who do not follow a piggybacking publication strategy.

As the general movement to increase publication productivity involves more players (regardless of their quality), we can expect an overall increase in citations (i.e., due to the increasing visibility of research in contemporary search engines). Such increasing or at least non-decreasing pattern in citations is going to be more in agreement with BHS. Some marginal unfair practices are unlikely to change the picture on the aggregate level.

The regional level of aggregation is believed to bring new insights into the problem of knowledge production, at least, in two ways. Formally, the regional level has a good statistical base describing the general economic environment, in which research institutions are *embedded*. If one uses a model on an institutional (organizational) level, the regional context can usually be captured by dummy variables mixing together many different spatial effects.

At the conceptual level, a region consists of territorial ties between *heterogeneous* research institutions located within the same area; these ties are known as agglomeration and (or) cluster effects (Gareev, 2012). In practical terms, they can manifest themselves in research facilities shared by R&D staff from different regional institutions or local university-industry networking.

Publication quality issues have been discussed in many sources as part of research policy discussions (Abramo et al., 2010; Butler, 2003). One of our goals is to inform policymakers focusing on quantitative measures, which can be shortsighted if we ignore the interplay between quality and quantity in research publication production.

So**, the policy advice** derived from our findings is relatively straightforward. Opting for purely quantitative performance indicators of publication activity can be a good strategy only at the very early stages of the convergence path. After gaining some initial experience with internationalization, policymakers should shift their focus to the quality of publications and provide additional support to research institutions that demonstrate higher research quality. Overall, higher quality is usually costly, and our model demonstrates that R&D expenses per employee can be converted into research quality, keeping volumetric scores unchanged.

Another point is the duality of quality indicators. To target quality, decision-makers should focus on both quality dimensions: relating to journal prestige and article citation impact. In practical terms, the quartile of a journal is a more robust and simple quality indicator that should be targeted in the first place. Citation impact indicators are more applicable for strategic evaluation in longer periods because they are more volatile in the short-run and harder to target.

The risks of ignoring quality dimensions may manifest themselves in the chase for publication scores, which in its turn may lead to quality degradation and shift the focus from efficiency to intensity.

**Open questions and directions for future exploration**

The quality versus quantity dilemma is often a policy concern in developing countries that are far from their steady state level of knowledge production. The regional level of the data aggregation adds some layers of complexity stemming from the need to account for potential sources of spatial and intertemporal lags.

Our approach is general enough to be adopted by many geostrategic entities for testing the ideas presented in this paper. However, numerous questions regarding unobserved heterogeneity and endogeneity in knowledge production models remain unanswered. One particular aspect that requires further exploration is the production possibilities frontier, where publication activity represents just a single facet among the numerous fields that can be readily quantified. A possible extension of the model involves relaxing certain assumptions in a theoretically grounded manner, which would enable the use of recently proposed estimation procedures based on maximum likelihood (see Leszczensky & Wolbring, 2019), or the adoption of hierarchical approaches. Additionally, a conventional approach to expanding the model is to include a broader set of controls, including new instrumental variables.

We would like to emphasize that our examination of the quality-impact relationship is conducted at the aggregated level, where we consider the overall framework rather than focus on individual authors. However, it would be intriguing to further investigate the disambiguation of publication records of individual authors and explore the distribution of researchers based on their impact. This analysis would provide insights into the contributions of new entrants and continuing researchers to high-quality research. While this topic is left for future exploration, it holds significant potential for further investigation.

Overall, we see the problem of dynamic publication quality accounts and thematic spatial lags in knowledge productivity measurement as a longstanding research interest.

**Declaration of competing interest**

The authors declare that they have no known competing financial interests or personal relationships that could have influenced the work reported in this paper.

**References**

1. Abramo, G., D'Angelo, C. A., & Di Costa, F. (2010). Testing the trade-off between productivity and quality in research activities. *Journal of the American Society for Information Science and Technology*, 61(1), 132-140. https://doi.org/10.1002/asi.21254

2. Abramo, G., D'Angelo, & C. A., Grilli, L. (2021). The effects of citation-based research evaluation schemes on self-citation behavior. *Journal of Informetrics,* 15(4). https://doi.org/10.1016/j.joi.2021.101204

3. Ali, M. A. (2021). Modeling regional innovation in Egyptian governorates: Regional knowledge production function approach. *Regional Science Policy and Practice*, 1-21. https://doi.org/10.1111/rsp3.12450

4. Baccini, A., De Nicolao, G., & Petrovich, E. (2019). Citation gaming induced by bibliometric evaluation: A country-level comparative analysis. *PLoS ONE,* 14(9), e0221212. https://doi.org/10.1371/journal.pone.0221212

5. Baltagi, B. (Ed.). (2015). *The Oxford Handbook of Panel Data*. Oxford University Press. https://doi.org/10.1093/oxfordhb/9780199940042.001.0001

6. Bautista-Puig, N., Moreno Lorente, L., & Sanz-Casado, E. (2021). Proposed methodology for measuring the effectiveness of policies designed to further research. *Research Evaluation,* 30(2), 215-229. https://doi.org/10.1093/reseval/rvaa021

7. Bettencourt, L. M. A., Kaiser, D. I., Kaur, J., Castillo-Chávez, C., & Wojick, D. E. (2008). Population modeling of the emergence and development of scientific fields. *Scientometrics,* 75(3), 495-518. https://doi.org/10.1007/s11192-007-1888-4

8. Biagioli, M. (2016). Watch out for cheats in citation game. *Nature,* 535(7611), 201. https://doi.org/10.1038/535201a

9. Binswanger, M. (2015). How nonsense became excellence: Forcing professors to publish. In I. Welpe, J. Wollersheim, S. Ringelhan, & M. Osterloh (Eds.), *Incentives and Performance: Governance of Research Organizations*. Springer, Cham, 19-32. https://doi.org/10.1007/978-3-319-09785-5_2

10. Bornmann, L. (2019). Does the normalized citation impact of universities profit from certain properties of their published documents - Such as the number of authors and the impact factor of the publishing journals? A multilevel modeling approach. *Journal of Informetrics,* 13(1), 170-184. https://doi.org/10.1016/j.joi.2018.12.007

11. Braun, T., Glänzel, W., & Schubert, A. (1989). Assessing assessments of British science. Some facts and figures to accept or decline. *Scientometrics*, 15, 165–170. https://doi.org/10.1007/BF02017195

12. Butler, L. (2003). Explaining Australia's increased share of ISI publications - The effects of a funding formula based on publication counts. *Research Policy,* 32(1), 143-155. https://doi.org/10.1016/S0048-7333(02)00007-0

13. Candia, C., Jara-Figueroa, C., Rodriguez-Sickert, C., Barabási, A., & Hidalgo, C. A. (2019). The universal decay of collective memory and attention. *Nature Human Behaviour,* 3(1), 82-91. https://doi.org/10.1038/s41562-018-0474-5

14. Charlot, S., Crescenzi, R., & Musolesi, A. (2015). Econometric modelling of the regional knowledge production function in Europe. *Journal of Economic Geography,* 15(6), 1227-1259. https://doi.org/10.1093/jeg/lbu035

15. Chen, K., Kou, M., & Fu, X. (2018). Evaluation of multi-period regional R&D efficiency: An application of dynamic DEA to China's regional R&D systems. *Omega (United Kingdom),* 74, 103-114. https://doi.org/10.1016/j.omega.2017.01.010

16. Cho, C.-C., Hu, M.-W., & Liu, M.-C. (2010). Improvements in productivity based on co-authorship: A case study of published articles in China. *Scientometrics,* 85(2), 463-470. https://doi.org/10.1007/s11192-010-0263-z

17. Coelli, T. J., Prasada Rao, D. S., O'Donnell, C. J., & Battese, G. E. (2005). *An introduction to efficiency and productivity analysis*. Springer, Boston, MA. https://doi.org/10.1007/b136381

18. Crescenzi, R., & Jaax, A. (2017). Innovation in Russia: The Territorial Dimension. *Economic Geography,* 93(1), 66-88. https://doi.org/10.1080/00130095.2016.1208532

19. Croissant, Y., & Millo, G. (2008). Panel data econometrics in R: The plm package. *Journal of Statistical Software,* 27(2), 1-43. https://doi.org/10.18637/jss.v027.i02

20. de Matos, C. M., Gonçalves, E., & Freguglia, R. D. S. (2021). Knowledge diffusion channels in Brazil: The effect of inventor mobility and inventive collaboration on regional invention. *Growth and Change,* 52(2), 909-932. https://doi.org/10.1111/grow.12467

21. Elsevier. (2020). *What is Field-weighted Citation Impact (FWCI)?* Retrieved February 14, 2022, from https://service.elsevier.com/app/answers/detail/a_id/14894/supporthub/scopus/~/what-is-field-weighted-citation-impact-%28fwci%29%3F/

22. Fritsch, M. (2002). Measuring the quality of regional innovation systems: A knowledge production function approach. *International Regional Science Review,* 25(1), 86-101. https://doi.org/10.1177/016001702762039394

23. Fritsch, M., & Slavtchev, V. (2007). Universities and innovation in space. *Industry and Innovation,* 14(2), 201-218. https://doi.org/10.1080/13662710701253466

24. Fritsch, M., & Slavtchev, V. (2010). How does industry specialization affect the efficiency of regional innovation systems? *Annals of Regional Science,* 45(1), 87-108. https://doi.org/10.1007/s00168-009-0292-9

25. Gareev, T. (2012). Clusters in the institutional perspective: on the theory and methodology of local socioeconomic development. *Baltic Region,* 3, 4–24. https://doi.org/10.5922/2079-8555-2012-3-1

26. Griliches, Z. (1979). Issues in Assessing the Contribution of Research and Development to Productivity Growth. *The Bell Journal of Economics,* 10(1), 92–116. https://doi.org/10.2307/3003321

27. Griliches, Z. (1990). Patent Statistics as Economic Indicators: A Survey. *Journal of Economic Literature,* 28(4), 1661-1707.

28. Halleck Vega, S., & Elhorst, J. P. (2015). The slx model. *Journal of Regional Science,* 55(3), 339-363. https://doi.org/10.1111/jors.12188

29. Haslam, N., & Laham, S. M. (2010). Quality, quantity, and impact in academic publication. *European Journal of Social Psychology,* 40(2), 216-220. https://doi.org/10.1002/ejsp.727

30. Hausman, J., Hall, B., & Griliches, Z. (1984). Econometric Models for Count Data with an Application to the Patents-R & D Relationship. *Econometrica,* 52, 909–938. https://doi.org/10.2307/1911191

31. Kim, Y. K., & Lee, K. (2015). Different impacts of scientific and technological knowledge on economic growth: Contrasting science and technology policy in East Asia and Latin America. *Asian Economic Policy Review,* 10(1), 43-66. https://doi.org/10.1111/aepr.12081

32. Kolesnikov, S., Fukumoto, E., & Bozeman, B. (2018). Researchers' risk-smoothing publication strategies: Is productivity the enemy of impact? *Scientometrics,* 116(3), 1995-2017. https://doi.org/10.1007/s11192-018-2793-8

33. Larivière, V., & Costas, R. (2016). How many is too many? On the relationship between research productivity and impact. *PLoS ONE,* 11(9), e0162709. https://doi.org/10.1371/journal.pone.0162709

34. Lawani, S. M. (1986). Some bibliometric correlates of quality in scientific research. *Scientometrics,* 9(1-2), 13-25. https://doi.org/10.1007/BF02016604

35. Leszczensky, L., & Wolbring, T. (2019). How to Deal With Reverse Causality Using Panel Data? Recommendations for Researchers Based on a Simulation Study. *Sociological Methods and Research*, 51(2), 837-865. https://doi.org/10.1177/0049124119882473

36. Levinsohn, J., & Petrin, A. (2003). Estimating production functions using inputs to control for unobservables. *Review of Economic Studies,* 70(2), 317-341. https://doi.org/10.1111/1467-937X.00246

37. Lindsey, D. (1989). Using citation counts as a measure of quality in science measuring what's measurable rather than what's valid. *Scientometrics,* 15(3-4), 189-203. https://doi.org/10.1007/BF02017198

38. Marrocu, E., Paci, R., & Usai, S. (2013). Proximity, networking and knowledge production in Europe: What lessons for innovation policy? *Technological Forecasting and Social Change,* 80(8), 1484-1498. https://doi.org/10.1016/j.techfore.2013.03.004

39. Martin, B. R. (2017). When social scientists disagree: Comments on the Butler-van den Besselaar debate. *Journal of Informetrics,* 11(3), 937-940. https://doi.org/10.1016/j.joi.2017.05.021

40. Mathies, C., Kivistö, J., & Birnbaum, M. (2020). Following the money? Performance-based funding and the changing publication patterns of Finnish academics. *Higher Education,* 79(1), 21-37. https://doi.org/10.1007/s10734-019-00394-4

41. Michalska-Smith, M. J., & Allesina, S. (2017). And, not or: Quality, quantity in scientific publishing. *PLoS ONE,* 12(6), e0178074. https://doi.org/10.1371/journal.pone.0178074

42. Miguélez, E., & Moreno, R. (2015). Knowledge flows and the absorptive capacity of regions. *Research Policy,* 44(4), 833-848. https://doi.org/10.1016/j.respol.2015.01.016

43. Millo, G. (2017). Robust Standard Error Estimators for Panel Models: A Unifying Approach. *Journal of Statistical Software,* 82(3). https://doi.org/10.18637/jss.v082.i03

44. Moed, H. F., Glänzel, W., & Schmoch, U. (2005). *Handbook of Quantitative Science and Technology Research. The Use of Publication and Patent Statistics in Studies of S&T Systems*. Dordrecht: Springer. https://doi.org/10.1007/1-4020-2755-9

45. Óhuallacháin, B., & Leslie, T. F. (2007). Rethinking the regional knowledge production function. *Journal of Economic Geography,* 7(6), 737-752. https://doi.org/10.1093/jeg/lbm027

46. Pakes, A., & Griliches, Z. (1980). Patents and R&D at the firm level: A first report. *Economics Letters,* 5(4), 377-381. https://doi.org/10.1016/0165-1765(80)90136-6

47. Perret, J. K. (2019). Re-Evaluating the Knowledge Production Function for the Regions of the Russian Federation. *Journal of the Knowledge Economy,* 10(2), 670-694. https://doi.org/10.1007/s13132-017-0475-z

48. Qin, X., & Du, D. (2019). A comparative study of the effects of internal and external technology spillovers on the quality of innovative outputs in China: The perspective of multistage innovation. *International Journal of Technology Management,* 80(3-4), 266-291. https://doi.org/10.1504/IJTM.2019.100287

49. Ramani, S. V., El-Aroui, M.-A., & Carrère, M. (2008). On estimating a knowledge production function at the firm and sector level using patent statistics. *Research Policy,* 37(9), 1568-1578. https://doi.org/10.1016/j.respol.2008.06.009

50. Rospatent. (2021). *Annual reports*. Retrieved February 9, 2022, from https://rospatent.gov.ru/ru/about/reports

51. Rosstat. (2021). *Regions of Russia. Socio-economic indicators*. Retrieved February 9, 2022, from https://rosstat.gov.ru/folder/210/document/13204

52. Sandström, U., & Van Besselaar, P. D. (2016). Quantity and/or quality? The importance of publishing many papers. *PLoS ONE,* 11(11), e0166149. https://doi.org/10.1371/journal.pone.0166149

53. Sebestyén, T., & Varga, A. (2013). Research productivity and the quality of interregional knowledge networks. *Annals of Regional Science,* 51(1), 155-189. https://doi.org/10.1007/s00168-012-0545-x

54. Seglen, P. O. (1997). Citations and journal impact factors: Questionable indicators of research quality. *Allergy: European Journal of Allergy and Clinical Immunology,* 52(11), 1050-1056. https://doi.org/10.1111/j.1398-9995.1997.tb00175.x

55. Sivertsen, G., Rousseau, R, & Zhang, L. (2019). Measuring scientific contributions with modified fractional counting. *Journal of Informetrics,* 13(2), pp. 679-694. https://doi.org/10.1016/j.joi.2019.03.010

56. Sutherland, W. J., Goulson, D., Potts, S. G., & Dicks, L. V. (2011). Quantifying the impact and relevance of scientific research. *PLoS ONE,* 6(11), e27537. https://doi.org/10.1371/journal.pone.0027537

57. Vadia, R., & Blankart, K. E. (2021). Regional innovation systems of medical technology: A knowledge production function of cardiovascular research and funding in Europe. *Region,* 8(2), 57-81. https://doi.org/10.18335/region.v8i2.352

58. Van den Besselaar, P., Heyman, U., & Sandström, U. (2017). Perverse effects of output-based research funding? Butler's Australian case revisited. *Journal of Informetrics,* 11(3), 905-918. https://doi.org/10.1016/j.joi.2017.05.016

59. Varun Shrivats, S., & Bhattacharya, S. (2014). Forecasting the trend of international scientific collaboration. *Scientometrics,* 101(3), 1941-1954. https://doi.org/10.1007/s11192-014-1364-x

60. Zemtsov, S., & Kotsemir, M. (2019). An assessment of regional innovation system efficiency in Russia: the application of the DEA approach. *Scientometrics,* 120(2), 375-404. https://doi.org/10.1007/s11192-019-03130-y

61. Zhou, P., Thijs, B., & Glänzel, W. (2009). Regional analysis on Chinese scientific output. *Scientometrics,* 81(3), 839-857. https://doi.org/10.1007/s11192-008-2255-9

**Table 4** Model for Main dataset, 2009-2021 (N = 936), **robust standard errors**

| log(PUB21EMP) | (1) | (2) | (3) | (4) | (5) | (6) | (7) |
|---|---|---|---|---|---|---|---|
| | **ols.q** | **fe.tw** | **fe.ow.q** | **fe.tw.q** | **fe.tw.q.sl.non** | **fe.tw.q.sl.noq** | **fe.tw.q.sl** |
| log(EXPEMP10) | 0.237 | 0.481*** | 0.565*** | 0.472*** | 0.472*** | 0.445*** | 0.460*** |
| | (0.191) | (0.082) | (0.130) | (0.083) | (0.082) | (0.080) | (0.082) |
| log(GRPCAP10) | 0.289* | 0.302 | 2.834*** | 0.306 | 0.334 | 0.299 | 0.323 |
| | (0.169) | (0.303) | (0.251) | (0.284) | (0.288) | (0.297) | (0.289) |
| log(PAPEMP) | 0.404*** | 0.246*** | 0.050 | 0.237*** | 0.247*** | 0.241*** | 0.242*** |
| | (0.083) | (0.072) | (0.083) | (0.067) | (0.071) | (0.068) | (0.069) |
| FWCI | 1.168*** | | 0.973*** | 0.346*** | 0.358*** | 0.299*** | 0.348*** |
| | (0.219) | | (0.168) | (0.091) | (0.091) | (0.088) | (0.089) |
| I(FWCI2) | -0.165*** | | -0.131*** | -0.051*** | -0.051*** | -0.043*** | -0.049*** |
| | (0.050) | | (0.033) | (0.015) | (0.015) | (0.015) | (0.015) |
| Q1SH | -0.010 | | -0.016*** | -0.008* | -0.010** | | -0.009** |
| | (0.009) | | (0.004) | (0.004) | (0.004) | | (0.004) |
| NQSH | -0.002 | | -0.001 | 0.003 | | 0.004** | 0.003 |
| | (0.005) | | (0.002) | (0.002) | | (0.002) | (0.002) |
| slFWCI | | | | | 1.433* | 1.291* | 1.569** |
| | | | | | (0.801) | (0.736) | (0.739) |
| slQ1SH | | | | | -0.027*** | | -0.021** |
| | | | | | (0.009) | | (0.009) |
| slNQSH | | | | | | 0.012** | 0.010* |
| | | | | | | (0.005) | (0.006) |
| factor(year)2010 | | 0.141*** | | 0.120*** | -0.038 | 0.019 | -0.052 |
| | | (0.028) | | (0.029) | (0.093) | (0.081) | (0.092) |
| factor(year)2011 | | 0.093 | | 0.055 | -0.062 | -0.106 | -0.128 |
| | | (0.064) | | (0.064) | (0.102) | (0.112) | (0.107) |
| factor(year)2012 | | 0.243*** | | 0.200** | 0.012 | -0.014 | -0.058 |
| | | (0.080) | | (0.082) | (0.149) | (0.155) | (0.150) |
| factor(year)2013 | | 0.420*** | | 0.341*** | 0.096 | 0.016 | -0.039 |
| | | (0.090) | | (0.091) | (0.182) | (0.198) | (0.189) |
| factor(year)2014 | | 0.702*** | | 0.603*** | 0.226 | 0.166 | 0.082 |
| | | (0.082) | | (0.086) | (0.236) | (0.251) | (0.243) |
| factor(year)2015 | | 0.958*** | | 0.843*** | 0.406 | 0.355 | 0.264 |
| | | (0.082) | | (0.084) | (0.277) | (0.284) | (0.275) |
| factor(year)2016 | | 1.090*** | | 0.957*** | 0.481 | 0.420 | 0.300 |
| | | (0.106) | | (0.111) | (0.296) | (0.306) | (0.304) |
| factor(year)2017 | | 1.258*** | | 1.141*** | 0.628** | 0.667** | 0.537* |
| | | (0.116) | | (0.117) | (0.318) | (0.303) | (0.305) |
| factor(year)2018 | | 1.451*** | | 1.326*** | 0.793** | 0.833*** | 0.685** |
| | | (0.137) | | (0.137) | (0.331) | (0.319) | (0.322) |
| factor(year)2019 | | 1.547*** | | 1.429*** | 0.849** | 0.963*** | 0.802** |
| | | (0.147) | | (0.146) | (0.349) | (0.323) | (0.329) |
| factor(year)2020 | | 1.606*** | | 1.518*** | 0.876** | 0.944** | 0.808** |
| | | (0.141) | | (0.138) | (0.392) | (0.366) | (0.366) |
| Constant | -4.724** | | | | | | |
| | (2.116) | | | | | | |
| Observations | 936 | 936 | 936 | 936 | 936 | 936 | 936 |

Note: *p<0.10; **p<0.05; ***p<0.01.

**ANNEX I – Recent contributions on empirical RKPF**

| Paper | Output / depended variable | Type of model and estimation strategy | Lagged variables | Explanatory variables | Dataset |
|---|---|---|---|---|---|
| Cho et al. (2010) | Total number of SCI published articles per researcher in the region | The Cobb-Douglas functional form of production functions based on the Jaffe Griliches–Acs Conceptual Model | A one-year time lag between the inputs and the outputs | (1) total per capita intramural expenditure on higher education in science and technology in the region;<br>(2) per capita research intensity in the region, calculated as the total number of researchers in the region divided by the region's total area;<br>(3) the total number of internationally published articles in the Science Citation Index;<br>(4) the total per capita domestic co-authorship capital, which is the total number of articles co-authored by researchers within the same region or in other regions;<br>(5) the total per capita international co-authorship capital, which is the total number of articles co-authored in the region with researchers from other countries. | A cross-sectional dataset of Chinese regions, 1998–2007 |
| Fritsch & Slavtchev (2010) | Number of disclosed patent applications in the region | The Cobb-Douglas form of a production function with a robust ordinary least squares (OLS) cross-section regression technique | A one-year time lag between the inputs and the outputs | (1) Number of private sector R&D employees in the region;<br>(2) Efficiency of RIS;<br>(3) Share of private sector R&D employees in the region;<br>(4) Universities' third-party funds from private companies per professor in the region;<br>(5) Average firm size in the region;<br>(6) Population density in the region;<br>(7) Share of regional employment in services;<br>(8) Share of employment in electrical engineering in the region;<br>(9) Regional index of industrial diversity. | A cross-sectional dataset of 97 German planning regions, 1996–2000 |
| Marrocu et al. (2013) | Total patents per million population | The general form of the empirical model for the Knowledge production function (KPF) is specified according to a log-linearized Cobb–Douglas production function and using | A three-year time lag between the inputs and the outputs. The explanatory variables are averaged over the three-year period to | (1) Total intramural R&D expenditure, over GDP;<br>(2) Population aged 15 and over with tertiary education, over total population;<br>(3) Population per $km^2$, thousands;<br>(4) Manufacturing employment over total employment; | A cross-sectional dataset of 276 regions in 29 European countries, 1999–2007 |

| | | OLS estimates | smooth away undue business cycle effects. | (5) Settlement structure typology. | |
|---|---|---|---|---|---|
| Sebestyén & Varga (2013) | The number of patent applications by region of the inventor; The number of publications in scientific journals in the ISI database. | The KPF specified by Romer and parameterized by Jones | A two-year time lag | (1) Gross regional expenditures on R&D; (2) Knowledge potential: the directly available knowledge from a region's partners; (3) Local density: the average number of links in a region's neighbourhood; (4) Global embeddedness: the structure of the network behind a region's immediate neighbourhood; (5) Ego network quality: a comprehensive measure of network position; (6) The number of a region's direct partners in the network; (7) National patent stock corresponding to the given region; (8) National publication stock corresponding to the given region; (9) Index of agglomeration. | A cross-sectional dataset of 189 European regions (a mix of NUTS2 and NUTS1 regions), 1998–2002 |
| Charlot et al. (2015) | The number of patents per million inhabitants | Generalized additive model (GAM) with log-log specification (Cobb–Douglas production function). Estimation of RKPF based on a semiparametric version of the random growth model. | Spatially lagged variables are computed by means of inverse Euclidian distance matrices | (1) GDP spent on R&D by public and private institutions, regardless of the Sector; (2) Regional share of workers with tertiary education or higher education; (3) Unobservable factors that influence regional innovative performance. | A panel data model for the entire EU-25, 1995–2004 |
| Kim & Lee (2015) | The ratio of corporate patent applications per million people in country i at time t; The ratio of SCI journal articles per million people. | Griliches' KPF model modifying by including scientific knowledge as an additional input. The OLS regression and fixed effect estimators were applied. | A one-year time lag for journal and corporate patent intensity | (1) The ratio of corporate patent applications per million people in country i at time t; (2) The ratio of SCI journal articles per million people; (3) The ratio of corporate R&D expenditure to GDP; (4) The number of persons who hold doctoral degrees in science and engineering from US universities (per million people); | A panel data model for East Asia and Latin America, 1960–2005 |

| | | | | (5) The ratio of exports to GDP. | |
|---|---|---|---|---|---|
| Miguélez & Moreno (2015) | The number of patent applications at the European Patent Office per million inhabitants using fractional counting | Baseline specification of regional KPF framework and estimating annual cross-section gravity models | A one-year time lag for the explanatory variables | (1) Regional R&D expenditures per capita;<br>(2) Regional endowments of human capital;<br>(3) Inward migration rate (IMR), i.e., the number of incoming inventors to region i over the number of local inventors in i, in a given time period t;<br>(4) The average number of co-inventions (co-patents) per inventor with inventors from outside the inventor's focal region. | A panel data model for 274 NUTS2 European regions, 2000–2007 |
| Crescenzi & Jaax (2017) | Patent applications counted according to the inventor's region of residence, per one million inhabitants | Augmented KPF | A two-period time lag between R&D investments and patent applications | (1) R&D expenditure as a percentage of regional GDP;<br>(2) Average of the R&D expenditure in neighbouring regions;<br>(3) Share of employees with higher education;<br>(4) Turnover of foreign enterprises as a percentage of regional GDP;<br>(5) Sectoral control variables. | A panel data model for 78 Russian regions, 1997-2011 |
| Qin and Du (2019) | - The total number of citations a paper receives within two years of being published;<br>- The number of invention patents per 10,000 population;<br>- The output value of new products in the high-tech industry. | The Cobb-Douglas functional form of production functions | A one-year time lag between the inputs and the outputs | (1) Foreign direct investment (FDI) stock;<br>(2) R&D expenditure of institute-university Projects;<br>(3) R&D expenditure of industry-university<br>(4) Projects;<br>(5) R&D expenditure of industry-institute Projects;<br>(6) Control variables. | A panel data model for 30 provinces in China, 2006–2015 |
| Ali (2021) | The number of patent applications per year | RKPF model with Exploratory spatial data analysis (ESDA) and OLS regression applied | A one-year time lag between the inputs and the outputs | (1) Gross expenditure on research and development;<br>(2) Professional employment in R&D in the private sector;<br>(3) Number of business support service | A cross-sectional dataset for the 27 Egyptian governorates, 2018 |

| | | | | (4) Concentration of high-tech industries. employees; | |
|---|---|---|---|---|---|
| de Matos et al. (2021) | The number of patents in the region per 100,000 population | The RKPF was estimated through a spatial dynamic panel. Its specification has been modified by using matrices of weights estimated by the gravitational models for interregional co-patents and mobility. | A one-period time lag between the dependent variable and the variables of interest. Models contain spatial lags of the dependent variable | (1) The number of patents of all regions, except region i, per 100,000 population; (2) The regional expenditure on research and development of companies; (3) The regional research and development expenditures of universities; (4) The regional human capital (the proportion of the local population holding a full university degree). | A panel data model for Brazilian regions, 2000–2011 |
| Vadia & Blankart (2021) | The number of publications in the region per capita | Empirical KPF model and using the spatial form of equation considering spatial proximity | Accounting for spatial lags of the dependent variable and the error term with robust Lagrange multiplier tests | (1) The economic performance by considering the GDP per region; (2) Innovatory effort - the regional level research investments as funding received by the EU Framework 7 and EU Horizon 2020 programmes. | A cross-sectional dataset of the 31 countries in Europe, 2014–2017 |

**Model for Articles and Reviews dataset, 2009-2021 (N = 936)**

(Standard and robust standard errors; p-values provided for standard errors)

| log(PUB21EMPA) | (1)         | (2)         | (3)         | (4)         | (5)              | (6)              | (7)          |
|---|---|---|---|---|---|---|---|
|                | ols.q.a     | fe.tw.a     | fe.ow.q.a   | fe.tw.q.a   | fe.tw.q.sl.a.non | fe.tw.q.sl.a.noq | fe.tw.q.sl.a |
| log(EXPEMP10)  | 0.212**     | 0.363***    | 0.463***    | 0.380***    | 0.384***         | 0.369***         | 0.380***     |
|                | (0.085)     | (0.048)     | (0.064)     | (0.048)     | (0.048)          | (0.048)          | (0.048)      |
|                | (0.177)     | (0.079)     | (0.113)     | (0.077)     | (0.080)          | (0.075)          | (0.077)      |
| log(GRPCAP10)  | 0.333***    | 0.433***    | 2.556***    | 0.437***    | 0.449***         | 0.433***         | 0.449***     |
|                | (0.066)     | (0.142)     | (0.103)     | (0.139)     | (0.140)          | (0.140)          | (0.140)      |
|                | (0.163)     | (0.270)     | (0.208)     | (0.262)     | (0.265)          | (0.263)          | (0.265)      |
| log(PAPEMP)    | 0.419***    | 0.240***    | 0.070*      | 0.233***    | 0.241***         | 0.232***         | 0.235***     |
|                | (0.029)     | (0.030)     | (0.039)     | (0.030)     | (0.030)          | (0.030)          | (0.030)      |
|                | (0.087)     | (0.072)     | (0.078)     | (0.069)     | (0.073)          | (0.070)          | (0.071)      |
| FWCIA          | 0.537***    |             | 0.668***    | 0.227***    | 0.235***         | 0.202***         | 0.236***     |
|                | (0.142)     |             | (0.081)     | (0.064)     | (0.064)          | (0.061)          | (0.064)      |
|                | (0.184)     |             | (0.127)     | (0.074)     | (0.074)          | (0.074)          | (0.074)      |
| I(FWCIA2)      | -0.104***   |             | -0.097***   | -0.041***   | -0.041***        | -0.038***        | -0.041***    |
|                | (0.022)     |             | (0.012)     | (0.009)     | (0.009)          | (0.009)          | (0.009)      |
|                | (0.022)     |             | (0.015)     | (0.009)     | (0.009)          | (0.009)          | (0.009)      |
| Q1SHA          | 0.010***    |             | 0.001       | -0.003      | -0.004**         |                  | -0.004       |
|                | (0.004)     |             | (0.003)     | (0.002)     | (0.002)          |                  | (0.002)      |
|                | (0.009)     |             | (0.004)     | (0.004)     | (0.004)          |                  | (0.004)      |
| NQSHA          | 0.014***    |             | 0.006***    | 0.004***    |                  | 0.004***         | 0.004***     |
|                | (0.003)     |             | (0.002)     | (0.001)     |                  | (0.001)          | (0.001)      |
|                | (0.005)     |             | (0.004)     | (0.003)     |                  | (0.003)          | (0.003)      |
| slFWCIA        |             |             |             |             | 0.166            | 0.212            | 0.466        |
|                |             |             |             |             | (0.539)          | (0.539)          | (0.553)      |
|                |             |             |             |             | (0.406)          | (0.579)          | (0.554)      |
| slQ1SHA        |             |             |             |             | -0.019*          |                  | -0.016       |
|                |             |             |             |             | (0.011)          |                  | (0.011)      |
|                |             |             |             |             | (0.005)          |                  | (0.006)      |
| slNQSHA        |             |             |             |             |                  | 0.003            | 0.001        |
|                |             |             |             |             |                  | (0.010)          | (0.010)      |
|                |             |             |             |             |                  | (0.005)          | (0.005)      |
| Observations   | 936         | 936         | 936         | 936         | 936              | 936              | 936          |

Note: time-dummies are omitted. Significance levels: *p<0.10; **p<0.05; ***p<0.01.

**Model for the Articles and Reviews dataset, 2009-2020 (N = 858)**

(Standard and robust standard errors: p-values provided for standard errors)

| log(PUB20EMPA) | (1) | (2) | (3) | (4) | (5) | (6) | (7) |
|---|---|---|---|---|---|---|---|
| | ols.q.a | fe.tw.a | fe.ow.q.a | fe.tw.q.a | fe.tw.q.sl.a.non | fe.tw.q.sl.a.noq | fe.tw.q.sl.a |
| log(EXPEMP10) | 0.216** | 0.353*** | 0.421*** | 0.377*** | 0.387*** | 0.360*** | 0.384*** |
| | (0.087) | (0.051) | (0.065) | (0.050) | (0.050) | (0.050) | (0.050) |
| | (0.179) | (0.077) | (0.106) | (0.077) | (0.078) | (0.074) | (0.076) |
| log(GRPCAP10) | 0.332*** | 0.559*** | 2.527*** | 0.551*** | 0.523*** | 0.553*** | 0.515*** |
| | (0.068) | (0.158) | (0.107) | (0.156) | (0.156) | (0.157) | (0.155) |
| | (0.168) | (0.291) | (0.183) | (0.278) | (0.280) | (0.280) | (0.276) |
| log(PAPEMP) | 0.407*** | 0.227*** | 0.062 | 0.220*** | 0.214*** | 0.218*** | 0.207*** |
| | (0.030) | (0.032) | (0.041) | (0.032) | (0.032) | (0.032) | (0.032) |
| | (0.087) | (0.072) | (0.074) | (0.068) | (0.068) | (0.070) | (0.066) |
| FWCIA | 0.662*** | | 0.707*** | 0.204*** | 0.217*** | 0.160** | 0.219*** |
| | (0.147) | | (0.083) | (0.067) | (0.067) | (0.064) | (0.067) |
| | (0.199) | | (0.131) | (0.077) | (0.077) | (0.077) | (0.077) |
| I(FWCIA2) | -0.115*** | | -0.099*** | -0.037*** | -0.037*** | -0.033*** | -0.037*** |
| | (0.022) | | (0.012) | (0.009) | (0.009) | (0.009) | (0.009) |
| | (0.023) | | (0.015) | (0.009) | (0.009) | (0.009) | (0.009) |
| Q1SHA | 0.005 | | -0.006** | -0.006** | -0.008*** | | -0.007*** |
| | (0.004) | | (0.003) | (0.002) | (0.002) | | (0.003) |
| | (0.009) | | (0.004) | (0.004) | (0.004) | | (0.004) |
| NQSHA | 0.014*** | | 0.006*** | 0.004*** | | 0.004*** | 0.004*** |
| | (0.003) | | (0.001) | (0.001) | | (0.001) | (0.001) |
| | (0.006) | | (0.004) | (0.003) | | (0.004) | (0.004) |
| slFWCIA | | | | | 0.511 | -0.023 | 0.931 |
| | | | | | (0.563) | (0.543) | (0.579) |
| | | | | | (0.319) | (0.672) | (0.540) |
| slQ1SHA | | | | | -0.111*** | | -0.119*** |
| | | | | | (0.031) | | (0.031) |
| | | | | | (0.036) | | (0.034) |
| slNQSHA | | | | | | 0.003 | 0.003 |
| | | | | | | (0.010) | (0.010) |
| | | | | | | (0.004) | (0.005) |
| Observations | 858 | 858 | 858 | 858 | 858 | 858 | 858 |

Note: time dummies omitted from the table. Significance levels for standard errors: *p<0.10; **p<0.05; ***p<0.01.